Letter

# Why even source-free gravity must be quantized

**S. Deser**[1,2,a]

[1] Walter Burke Institute for Theoretical Physics, California Institute of Technology, Pasadena, CA 91125, USA
[2] Physics Department, Brandeis University, Waltham, MA 02454, USA

Received: 20 April 2022 / Accepted: 29 April 2022 / Published online: 10 May 2022

**Abstract** I note, on the most elementary grounds: its classical ultraviolet catastrophe, that even source-free General Relativity must be quantized, despite some eminent opinions to the contrary and with a little fine print.

## 1 Quantization from ultraviolet catastrophe

It is almost universally agreed that General Relativity (GR) must be quantized, if only because its matter sources are, let alone on more speculative string theoretic or supergravity grounds. However, a few prominent doubters, whom I shall not cite, have claimed classical immunity for GR. One aim of this note is to provide the simplest and perhaps most credible reason for the necessity of GR's quantization (QG), even absent sources: the same ultraviolet catastrophe (UVC) that motivated Planck to introduce photons [1]. To be sure, one might object that the resultant QED is perturbatively renormalizable, whereas QG is not, so that removing this catastrophe opens GR to even more serious infinities. This argument, that the good is the enemy of the best is, however, irrelevant – one indeed expects modifications to GR for a viable quantization – but Planck's is a necessary first step under all scenarios. Also, it goes without saying that quantization, although forced by UV problems, applies for all frequencies as for Maxwell.

Since the UVC is taught to all undergraduates, I only mention its relevant aspect: it is independent of any dynamics, being just thermodynamic, leading to the rapid growth with frequency of the number of excited modes in a cavity, independent of the field species. So already weak gravitational waves – and we now know at least those exist – have exactly the same behavior as Maxwell's, given the identical number of helicities of all $s > 0$ massless fields. Hence Planck's solution is both necessary and sufficient: no calculations necessary. While GR is not merely its linearized limit, however, and the above thermodynamic counting is usually independent of a field's nonlinearities or interactions provided it has a normal kinetic term, as original and cosmological GR both do, still, the possibility that GR's self-attraction might remove the UVC must disposed of. The simple resolution is that instead of the Newtonian formula for total energy of a self-interacting system,

$$\mathrm{E} = E_0 - G{E_0}^2/r, \tag{1}$$

in GR it reads [2]

$$\mathrm{E} = E_0 - G\mathrm{E}_0^2/r, \tag{2}$$

where $E_0$ is the bare energy, and r a characteristic radius (all energy, including gravitational, interacts!). Solving the quadratic (2) for E reveals that it not only does not become arbitrarily negative, but rises, as $\sqrt{E_0}$, hence still becomes infinite as $E_0$ does, so the – slightly weaker – UVC remains. Note that, while (2) was derived in a particular frame, the value, as against the form, of E is an invariant; recall also the E is necessarily positive and vanishes only at source-free flat space. A last small detail: strictly, thermodynamics requires the notion of energy inside the cavity, but as is well-known, gravitational energy is only defined for a full system, not just inside a finite volume. However, the cavity's walls are infinitely high, so that's good enough for our purposes as a working energy notion, or like energy in our universe, dS GR [3]: inside its intrinsic horizon that surely confines all! Indeed, that is also a reasonable confined (inside the intrinsic horizon) thermo ensemble: time-dependent quadrupole matter is present, just like its charged dipole analog.

Also, one might ask whether or why the Planck constant here should be the same as for matter. That is easily disposed of both in theory and observationally as in the photo effect in ED, where the Planck constants of electrons and photons are clearly matched. Here one may always rescale Newton's constant to achieve this equality, again required by the interaction with matter.

[a] e-mail: deser@brandeis.edu (corresponding author)

**Acknowledgements** I thank Adam Schwimmer, without whom this note would not exist. This work was supported by the U.S. Department of Energy, Office of Science, Office of High Energy Physics, Award Number de-sc0011632.

**Data availability statement** This manuscript has no associated data or the data will not be deposited. [Authors' comment: There is no data associated with this article.]

Funded by SCOAP$^3$.

## References

1. M. Planck, Berliner Berichte 18 May 1899 *et seq.*
2. R.L. Arnowitt, S. Deser, C.W. Misner, Phys. Rev. Lett. **4**, 375 (1960)
3. L. Abbott, S. Deser, Nucl. Phys. B **95**, 76 (1982)